\documentclass[twocolumn,showpacs,amsmath,amssymb,floatfix]{revtex4}

\usepackage{graphicx}
\begin{document}

\title{ Phase diagram of the half-filled two-dimensional  SU(N) Hubbard-Heisenberg model: 
a quantum Monte Carlo study. }
\author{F.F. Assaad}
\affiliation{ Institut f\"ur Theoretische Physik und Astrophysik
Universit\"at W\"urzburg, Am Hubland D-97074 W\"urzburg }

\begin{abstract}
We investigate  the phase diagram of the 
half-filled $SU(N)$ Hubbard-Heisenberg model with hopping $t$, exchange $J$ and Hubbard $U$,   
on a two-dimensional square lattice.   
In the large-N limit, and as a function of decreasing values of $t/J$,  the model shows a transition 
from a d-density wave state to a spin dimerized 
insulator.  A similar behavior is observed at $N=6$ 
whereas at $N=2$ a spin density wave insulating ground state 
is stabilized.  The $N=4$ model, has  a d-density
wave ground state at large  values of $t/J$  which  as a function of decreasing values of $t/J$  
becomes unstable to an insulating state  with no apparent lattice and spin broken symmetries.  
In this state, the  staggered spin-spin correlations decay as a power-law, resulting in
gapless  spin excitations at  $\vec{q} = (\pi,\pi)$.  Furthermore, 
low lying spin modes with small spectral weight  are apparent around the wave vectors 
$\vec{q} = (0,\pi)$   and $\vec{q} = (\pi,0)$.  This gapless spin liquid state is equally found in
the $SU(4)$ Heisenberg ($ U/t \rightarrow \infty $ ) model  in the self-adjoint 
antisymmetric representation.  An interpretation of this state in terms of a  $\pi$-flux phase is 
offered.  Our  results  stem from projective ($T=0$) quantum Monte-Carlo simulations  
on lattice sizes  ranging up to $24 \times 24$.
\end{abstract}

\preprint{NSF-KITP-04-60}
\maketitle

\section{ Introduction }

$SU(N)$  symmetric models of correlated electron systems have attracted considerable interests 
in the  past decades.  For instance, those models are relevant for the understanding of Mott insulators  
with orbital degeneracy as described by the Kugel-Khomskii Hamiltonian \cite{Kugel82}. 
For two-fold orbital degeneracy 
and at a point where the orbital and spin degrees of freedom  play a very symmetric role, 
this model maps onto an $SU(4)$ symmetric Hubbard, or Heisenberg model with fundamental representation 
on each site \cite{Li88}.   It has also recently been argued that realizations of $SU(N)$ Hubbard models 
are at reach in the context of optical lattices \cite{Honerkamp04}.

$SU(N)$ generalizations of $SU(2)$ lattice fermion models can be solved 
exactly in the large-N limit.  Systematic corrections in terms of Gaussian fluctuations 
around the mean-field or saddle point solution may be computed.  The simplifications which 
occur in the large-$N$ limit, namely the suppression of quantum fluctuations have important 
consequences  for auxiliary field quantum Monte Carlo (QMC) simulations. 
As a function of growing values of $N$ the negative sign problem inherent to  stochastic methods 
is reduced thus rendering simulations more and more tractable.  In fact, some generalizations of 
Hubbard models lead to the absence of sign problems for specific values of $N$  and irrespective 
of doping \cite{Assaad02a,Wu03}. However, the extrapolation from the  soluble  large-N   limit to 
the physical $N=2$ case is by no means unambiguous since phase transitions can occur as a function of  
$N$. 

In this article,  we will primarily  concentrate  on the half-filled Hubbard-Heisenberg model 
on a square lattice 
and map out it's phase diagram as a  function of $N$ and coupling strength. 
At this band filling, the sign problem  is absent for even values of $N$.  Hence,  
ground state properties  can be investigated on  lattice sizes ranging up to $24 \times 24 $ unit cells. 
We will show  the existence  of d-density wave (DDW) states down to $N=4$ and  of spin-dimerized  states 
at  $N=6$.  The most intriguing result is a  possible realization of  a gapless spin-liquid 
phase  for the  $N=4$  model in the  Heisenberg limit.  

The $SU(N)$  symmetric Hubbard-Heisenberg model we consider  reads:
\begin{eqnarray}
 & & 	H  = H_t + H_U + H_J  \; \; \;  {\rm with } \nonumber  \\
         & &  H_t =  - t \sum_{ \langle \vec{i},\vec{j} \rangle }    \vec{c}^{\dagger}_{\vec{i}}
              \vec{c}_{\vec{j}} + {\rm H.c.}  \nonumber \\
	 & &  H_U =  \frac{U}{N}  \sum_{\vec{i}} \left( 
             \vec{c}^{\dagger}_{\vec{i}}  \vec{c}_{\vec{i}} - \rho {\frac{N}{2} } \right)^2
          \nonumber \\
& & 	H_J =  -\frac{J}{2 N}  \sum_{ \langle \vec{i},\vec{j} \rangle  } \left( 
           D^{\dagger}_{ \vec{i},\vec{j} } D_{ \vec{i},\vec{j}}  + 
           D^{}_{ \vec{i},\vec{j} } D^{\dagger}_{ \vec{i},\vec{j} }  \right) .
\end{eqnarray}
Here, 
$ \vec{c}^{\dagger}_{\vec{i}} = 
(c^{\dagger}_{\vec{i},1},  c^{\dagger}_{\vec{i},2}, \cdots, c^{\dagger}_{\vec{i}, N } ) $  is an
$N$-flavored spinor, $ D_{ \vec{i},\vec{j}} = \vec{c}^{\dagger}_{\vec{i}} 
\vec{c}_{\vec{j}}  $ and $\rho $ corresponds 
to the band-filling. 
At $N=2$, the  operator identity  
\begin{eqnarray}
\frac{-1}{4}  & & \left( D^{\dagger}_{ \vec{i},\vec{j} } D_{ \vec{i},\vec{j}}  + 
           D^{}_{ \vec{i},\vec{j} } D^{\dagger}_{ \vec{i},\vec{j} }  \right)  
= \nonumber \\  
& & \vec{S}_{\vec{i}} \cdot  \vec{S}_{\vec{j}}  + \frac{1}{4}  
       \left[ (n_{\vec{i}}-1) (n_{\vec{j}}-1)  - 1  \right]
\end{eqnarray}
holds. Here, the fermionic representation of the spin $1/2$ operator reads
 $ \vec{S}  = \frac{1}{2} \sum_{s,s'} c^{\dagger}_{s} \vec{\sigma}_{s,s'} c_{s'} $   where 
$ \vec{\sigma}$ are the Pauli spin matrices.  Thus, at $N=2$ the model reduces to the standard 
Hubbard-Heisenberg model. 

In the strong coupling limit, $U/t \rightarrow \infty $,  and  at integer values of $\rho N/2$, 
charge fluctuations are suppressed. The model maps onto the $SU(N)$  Heisenberg  Hamiltonian
\begin{equation}
	H = \frac{J}{N} \sum_{ \langle \vec{i}, \vec{j} \rangle } 
\sum_{\alpha,\beta} S_{\alpha,\beta,\vec{i}} S_{\beta,\alpha,\vec{j}}
\end{equation} 
with 
\begin{equation}
\label{generators}
  S_{\alpha,\beta,\vec{i}}  = c^{\dagger}_{\alpha,\vec{i}}  c_{\beta,\vec{i}}  - \frac{1}{N} 
   \delta_{\alpha,\beta} \sum_{\gamma}  c^{\dagger}_{\gamma,\vec{i}}  c_{\gamma,\vec{i}}
\end{equation} 
the generators of  $SU(N)$ satisfying the commutation relations:
\begin{equation}
 	 \left[  S_{\alpha,\beta,\vec{i}}, S_{\gamma,\delta,\vec{j}} \right] = 
\delta_{\vec{i},\vec{j}} \left( S_{\alpha,\delta,\vec{i}} \delta_{\gamma,\beta} - 
                                S_{\gamma,\beta,\vec{i} } \delta_{\alpha,\delta} \right ).
\end{equation}
The representation of the $SU(N)$ group is  determined by the local constraint
\begin{equation}
\label{Constraint}
	\vec{c}^{\dagger}_{\vec{i}} \vec{c}_{\vec{i}} = \rho \frac{N}{2}.
\end{equation}
In the terminology of Young tableaux the above leads to a  tableau with $\rho N/2$ rows and a single 
column. In particular,  at $N=4$, and $\rho = 1/2$ (quarter band-filling)  the model maps onto the 
$SU(4)$ symmetric Kugel-Khomskii Hamiltonian with fundamental representation of $SU(4)$ on each 
lattice site.  A study of large-N Heisenberg models in various representations may be found in Ref. 
\cite{Read89}.

$SU(N)$ Heisenberg models have been considered numerically in Refs. 
\cite{Santoro99,Harada03}. 
Those models differ substantially  from ours  in the choice of the representation.
On one sublattice the fundamental representation (Young tableau with one row and a single column) 
is considered and on the other the adjoint representation 
(Young tableau with $N-1$ rows and a single column).  Based on Green function Monte-Carlo methods, 
it has been argued this $SU(4)$ model  has a spin-liquid ground state. However, simulations on larger 
lattice sizes with the loop algorithm have shown that the model has a broken symmetry ground state 
\cite{Harada03}. In contrast, our results for the $SU(4)$  Heisenberg model, at $\rho = 1$  and in the
self-adjoint representation (see Eq. (\ref{Constraint})) point towards an insulating state 
with no broken symmetries.

The article is organized as follows. 
In the next section we formulate the  the partition function of  the model as a path integral  
over bosonic field. This formulation constitutes the starting point for both the saddle 
point approximation and the  auxiliary field quantum Monte Carlo simulations.
In Section \ref{Results} we  present the phase diagram of  
the half-filled model as a function of $N$ and coupling constants.  
Finally, we summarize and draw conclusions. 

\section {Large-N limit and quantum Monte-Carlo simulation.}
Both the saddle point approximation as well as the auxiliary field QMC rely on a 
path integral formulation of the partition function. 
Using the Trotter decomposition, we write the partition function as:
\begin{equation}
	Z = {\rm Tr} \left[ e^{- \beta H } \right] =   
	{\rm Tr } \left[ \prod_{n = 1}^{m} e^{-\Delta \tau H_t } e^{-\Delta \tau H_U }  
        e^{-\Delta \tau H_J } \right]  
\end{equation} 
Here $m \Delta \tau = \beta $ and we have omitted the 
systematic error of oder  $ \Delta \tau^2 $. 
Using the Hubbard Stratonovich (HS) transformation, we introduce bosonic fields to decouple 
the two body interaction terms. 
Let us start with the Heisenberg term which we write -- replacing the sum over nearest neighbors
$ \langle \vec{i}, \vec{j} \rangle $ by a sum over bonds $b$ -  in terms  of perfect squares 
\begin{equation}
 	H_J =  -\frac{J}{4 N}  \sum_{ b } 
        \left(D^{\dagger}_{ b } +  D_{ b }  \right)^2  - 
        \left(D^{\dagger}_{ b } -  D_{ b }  \right)^2.
\end{equation}
We can now apply the standard HS transformation to obtain: 
\begin{eqnarray}
	& & e^{-\Delta \tau H_J } \propto \int \prod_{ b } {\rm d } 
\gamma_{ b } {\rm d } \eta_{b }  \\
 & &  e^{ - \sum_b \left[ \frac{\gamma_{ b }^2 } {2} + \frac{\eta_{ b }^2 } {2} - \sqrt{\frac{ J \Delta \tau}{2 N} } 
\left( \gamma_b( D^{\dagger}_b + D_b) + i \eta_b( D^{\dagger}_b - D_b)    \right) \right] } \nonumber \\
& & \propto 
\int \prod_{ b } {\rm d Re} z_b  {\rm d Im} z_b  
e^{ - \sum_b \left[ N \Delta \tau J  | z_{ b }|^2    -  \Delta \tau J \left( z_b D_b^{\dagger} + 
     \bar{z}_b D_b \right) \right] }. \nonumber 
\end{eqnarray}
In the above, we  introduce a complex variable  per bond:  
$  z_b = ( \gamma_b + i \eta_b) / \sqrt{ 2N J \Delta \tau } $. 
Following the same steps, we decouple the Hubbard term as: 
\begin{equation}
	e^{-\Delta \tau H_U } \propto \int \prod_{\vec{i}} {\rm d} \Phi_{\vec{i}} 
   e^{- \sum_{\vec{i}} \left[ N \Delta \tau U  \Phi_{\vec{i}}^2/4 - 
   i \Delta \tau U \Phi_{\vec{i}} \left( \vec{c}^{\dagger}_{\vec{i}} \vec{c}_{\vec{i}} - \rho N/2 \right) 
   \right] }.
\end{equation}

Using the above transformations, the partition function in the limit 
$ \Delta \tau \rightarrow 0 $  is given by a functional integral over the  space and 
imaginary time dependent HS fields:
\begin{equation}
	Z \propto \int \prod_{\vec{i}} {\rm D }  \Phi_{\vec{i}} (\tau) 
         \prod_{ b } {\rm D Re} z_{b}(\tau)  {\rm D Im} z_{b}(\tau) 
     e^{- N S( \left\{ \Phi \right\} , 
 \left\{ z \right\} , \left\{ \bar{z}  \right\} ) }  .
\end{equation}
The action reads:
\begin{eqnarray}
	 S \left( \left\{ \Phi \right\} , \left\{ z \right\} , \left\{ \bar{z}  \right\} \right) & = &  
   \underbrace{\int {\rm d} \tau  J \sum_{b} | z_{ b } (\tau) |^2  + 
    U \sum_{\vec{i}} \Phi^2_{\vec{i}}(\tau)/4}_{=S_0} 
\nonumber \\ 
    & &  - \ln {\rm Tr}   \left[  T e^{-\int_{0}^{\beta} {\rm d} \tau h(\tau)  } \right].
\end{eqnarray}
with 
\begin{eqnarray}
      h(\tau) & & = -  \sum_{ \langle \vec{i},\vec{j} \rangle } 
 \left[ \left( t + J \bar{z}_{ \langle \vec{i},\vec{j} \rangle } (\tau)  \right) 
     c^{\dagger}_{\vec{i}}  c_{\vec{j}}  + {\rm H.c.} \right]  \nonumber \\
   & & - i U \sum_{\vec{i}} \Phi_{\vec{i}} (\tau) \left( c^{\dagger}_{\vec{i}} c_{\vec{i}} - \rho/2 \right) 
\end{eqnarray} 
Notice that in the above definition of $h(\tau)$, the creation and annihilation operators are not 
N-flavored spinors but correspond to spinless fermion operators. 

\subsection{The Saddle point.}

In the large-N limit, the saddle point approximation,
\begin{equation}
	\frac{ \delta S }{\delta z_b(\tau) } = 
	\frac{ \delta S }{\delta \bar{z}_b(\tau) } = 
	\frac{ \delta S }{\delta \Phi_{\vec{i}} (\tau) } =  0
\end{equation}  
becomes exact.  Assuming time independent fields,
we derive the mean-field equations: 
\begin{eqnarray}
        \Phi_{\vec{i}}  &  = &    i \left(    
              2 \langle c^{\dagger}_{\vec{i}} c_{\vec{i}}\rangle_{h}  - \rho  \right) \nonumber \\
        z_{\langle \vec{i}, \vec{j} \rangle }  & = & \langle c^{\dagger}_{\vec{i}} c_{\vec{j}} \rangle_h  
\nonumber \\
        \bar{z}_{\langle \vec{i}, \vec{j} \rangle }  & = &
    \langle c^{\dagger}_{\vec{j}} c_{\vec{i}} \rangle_h.
\end{eqnarray}

\begin{figure}[h]
\begin{center}
\includegraphics[width=.2\textwidth]{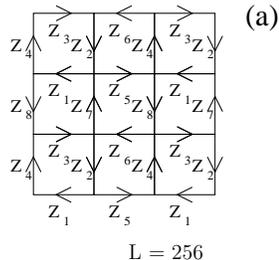} \\ 
\vspace{0.2cm}
\includegraphics[width=.45\textwidth]{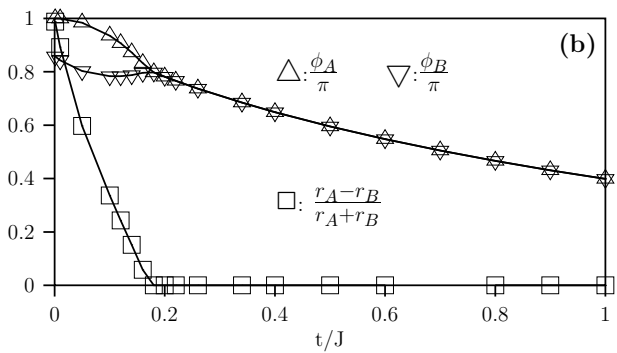} \\
\end{center}
\caption[]{ (a) The four site unit cell with lattice vectors, $\vec{a}_1 = 2 \vec{a}_x$ and 
$\vec{a}_2 = 2 \vec{a}_y$ and corresponding fields. Here, $\vec{a}_x$ and $\vec{a}_y$ correspond to the 
lattice vectors of the underlying square lattice. 
(b)   Mean-field order parameters (see Eq. \ref{MF_oder_parameters})  as obtained from the saddle point 
equations. }
\label{Mean.fig}
\end{figure}

\begin{figure}[h]
\begin{center}
\includegraphics[width=.4\textwidth]{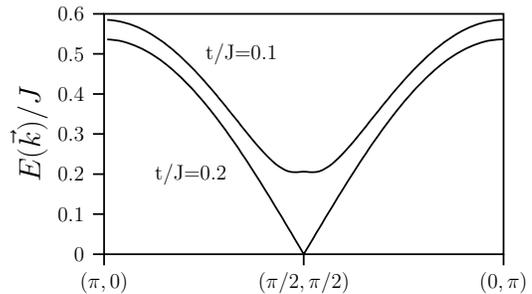} \\ 
\end{center}
\caption[]{ Lowest lying single particle excitation at half-band filling. 
The transition from the DDW state to dimerized  state corresponds to a semi-metal to insulator transition.  }
\label{Disp.fig}
\end{figure}

The above saddle point has been considered by Affleck and Marston \cite{Affleck88,Marston89}. 
At half-band filling, $\rho = 1$, and {\it large } values of $t/J$ 
a d-density wave state is realized. This state becomes unstable towards 
dimerization  as the coupling $t/J$ is reduced.  Here, we have solved the 
mean-field equations for a four site unit-cell (see Fig. \ref{Mean.fig}), thus allowing  
more freedom in the dimerization pattern than in  \cite{Marston89}. With 
\begin{equation}
	Z_i \equiv  r_i e^{i \phi_i/4},
\end{equation} 
the solutions we find are characterized by:
\begin{eqnarray}
\label{MF_oder_parameters}
	r_1 = r_2 = r_3 = r_4 = r_A,   
	r_5 = r_6 = r_7 = r_8 = r_B \nonumber \\ 
	\phi_1 = \phi_2 = \phi_3 = \phi_4 = \phi_A,   
	\phi_5 = \phi_6 = \phi_7 = \phi_8 = \phi_B. 
\end{eqnarray}
The values of  the oder parameters are plotted in Fig. \ref{Mean.fig}. As apparent the d-density wave state
with $r_A= r_B$ and $\phi_A = \phi_B $ is unstable towards box  dimerization below 
$t_c/J \simeq 0.17 $ \cite{Dombre89}. 
The DDW state is a semi-metal since gapless single particle excitations  are present at wave vectors 
$(\pm \pi/2a, \pm \pi/2a )$.   Dimerization opens a  quasiparticle gap at all wave vectors. Hence the 
transition from the DDW to the dimerized phase corresponds to a semi-metal to insulator transition as 
shown in Fig. \ref{Disp.fig}.   

We note that the results of Affleck and Marston \cite{Marston89} may be recovered by imposing: 
\begin{equation}
        Z_5 = Z_3,  Z_6 = Z_1,  Z_7 = Z_4, \; \;  {\rm and}   \; \;   Z_8 = Z_2. 
\end{equation}

\subsection{The Monte Carlo simulation.}
The Monte-Carlo  approach
relies on the same formulation of the partition function.  Before discussing 
details of the implementation  let us concentrate on our primary concern, namely the sign 
problem.   In general, $ e^{-NS(\phi, z, \bar{z} ) }$ is not a positive quantity and hence may 
not  be interpreted as an unnormalized probability distribution from which we  sample  
field configurations.  Hence, in the Monte Carlo method, we  consider the probability
distribution: 
\begin{equation}
	P( \phi, z, \bar{z} ) = \frac{ \left| e^{ - N S ( \phi,z,\bar{z} )  } \right| }  
{ \int {\cal D} \left[ \phi, z, \bar{z} \right]  \left| e^{ - N S ( \phi,z,\bar{z} )  } \right| }
\end{equation}
and estimate  the expectation value of an observable with: 
\begin{equation}
	\langle O \rangle =  
                     \frac{ \int {\cal D} \left[ \phi, z, \bar{z} \right] P( \phi, z, \bar{z} ) 
			e^{i \delta(\phi, z, \bar{z}) }  O( \phi, z, \bar{z}) }
                          { \int {\cal D} \left[ \phi, z, \bar{z} \right] P( \phi, z, \bar{z} ) 
			e^{i \delta(\phi, z, \bar{z}) } }.
\end{equation}
In the above, $  O( \phi, z, \bar{z}) $ is the  expectation value of the observable for a  given 
configuration of fields, and 
$ e^{ - N S ( \phi,z,\bar{z} )  } = 
\left| e^{ - N S ( \phi,z,\bar{z} )  } \right| e^{i \delta(\phi, z, \bar{z}) }$.
The denominator in the above equation, corresponds to the average sign: 
$ \langle e^{i \delta } \rangle_P $.

In the large-N limit, where the saddle point approximation becomes exact, the average sign is temperature
independent and equal to unity. On the other hand, it is known that for the SU(2) model the average sign 
decays as $e^{- \Delta V \beta }$ where $V$ corresponds to the volume of the system and $ \Delta $ is 
a positive constant. Hence, we can
conjecture that $\Delta $ is a decreasing function of $N$. This has for consequence that at a given
temperature the  sign problem becomes less and less severe as function of growing values of $N$. We have
checked this numerically for the quarter filled, $\rho=1/2$, model.    Unfortunately  the sign problem 
for the $SU(4)$ quarter-filled model -- the $SU(4)$ symmetric
Kugel-Khomskii model -- was still to severe to study the nature of the Mott insulating 
phase  in the strong coupling limit.

\begin{figure}[h]
\begin{center}
\includegraphics[width=.4\textwidth]{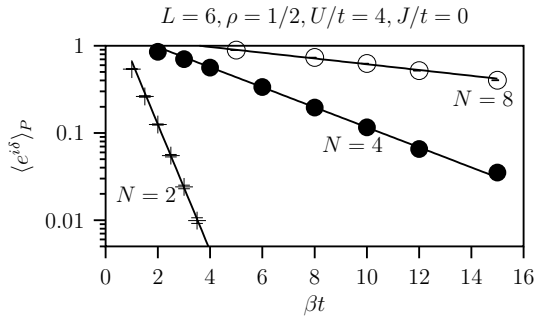} 
\end{center}
\caption[]{ Average sign as a function of $N$ for the quarter-filled $SU(N)$ Hubbard model. The  solid
line corresponds to a fit to the form: $ae^{-b\beta}$. }
\label{Sign.fig}
\end{figure}

At half-band filling, $\rho=1$, and even values of $N$,   particle-hole symmetry  leads to the absence of a 
minus-sign problem. At this filling,  and under the canonical transformation 
\begin{equation}
	c^{\dagger}_{\vec{i}} \rightarrow (-1)^{i_x + i_y}  c^{}_{\vec{i}}
\end{equation}
the Hamiltonian $h(\tau)$ transforms as: 
\begin{equation}
	h(\tau) \rightarrow \overline{ h(\tau) } 
\end{equation}
such that 
\begin{equation}
	{\rm Tr} \left[ T e^{-\int_{0}^{\beta} {\rm d} \tau h(\tau) } \right]   =
     \overline{	{\rm Tr} \left[ T e^{-\int_{0}^{\beta} {\rm d} \tau h(\tau) } \right]   }.
\end{equation}
Hence, the above quantity is real and 
\begin{equation}
	e^{-N S} = e^{-N S_0} 
\left[ { \rm Tr} \left[ T e^{-\int_{0}^{\beta} {\rm d} \tau  h(\tau) } \right]  \right]^{N} 
\end{equation}
is positive for even values of $N$.

We now summarize the technicalities required to carry  efficient simulations. 
Since we are interested in ground state  properties, it is more efficient to adopt a projective method 
based on the equation:
\begin{equation}
\label{PQMC}
	\langle O \rangle_0  = \lim_{\beta \rightarrow \infty} 
       \frac{ \langle \Psi_T | e^{-\beta H/2} O e^{-\beta H/2} | \Psi_T \rangle }
            { \langle \Psi_T | e^{-\beta H}  | \Psi_T \rangle }.
\end{equation} 
The trial wave function $ | \Psi_T \rangle $ is required to be non-orthogonal to the ground state and 
$\beta $  corresponds to a projection parameter.  For the trial wave function we choose the form: 
\begin{equation}
\label{Trial}
	| \Psi_T \rangle = | \Psi_T \rangle_1 \otimes | \Psi_T \rangle_2 \otimes \cdots 
       \otimes | \Psi_T \rangle_N 
\end{equation}
where $ | \Psi_T \rangle_\alpha $ is the ground state of the single particle Hamiltonian 
$ -t \sum_{\langle \vec{i}, \vec{j} \rangle } c^{\dagger}_{\vec{i},\alpha}  c_{\vec{j},\alpha}  + 
{\rm H.c.} $ in the  flavor $\alpha$ Hilbert space.  With this choice of  trial wave function, the 
action within  the projection formalism takes the form: 
\begin{equation}
	S = S_0 - \ln \langle \chi_T | T  e^{-\int_{0}^{\beta} {\rm d} \tau h(\tau) } | \chi_T \rangle 
\end{equation}
where  $ | \chi_T \rangle $ is the ground state of the spinless fermion Hamiltonian:
$ -t \sum_{\langle \vec{i}, \vec{j} \rangle } c^{\dagger}_{\vec{i}}  c_{\vec{j}}  + 
{\rm H.c.} $.   In the simulations we will present in the next section, we have typically 
used $ \beta J = 40 $ which we found to be sufficient to filter out the ground state from the 
trial wave function within statistical uncertainty.

We  use a finite imaginary time-step $\Delta \tau $ which we have set to 
$\Delta \tau J = 0.1$.  This introduces a systematic error of the order $\Delta \tau^2$. Given this 
systematic error, it is much more efficient to use an approximate  discrete HS transformation to decouple 
the perfect square term: 
\begin{equation}
        e^{\Delta \tau  \lambda  A^2 } = 
        \sum_{ l = \pm 1, \pm 2}  \gamma(l)
e^{ \sqrt{\Delta \tau \lambda }
       \eta(l)  A }
                + {\cal O} (\Delta \tau ^4)
\end{equation}
where the fields $\eta$ and $\gamma$ take the values:
\begin{eqnarray}
 \gamma(\pm 1) = 1 + \sqrt{6}/3, \; \; \gamma(\pm 2) = 1 - \sqrt{6}/3
\nonumber \\
 \eta(\pm 1 ) = \pm \sqrt{2 \left(3 - \sqrt{6} \right)},  \; \;
 \eta(\pm 2 ) = \pm \sqrt{2 \left(3 + \sqrt{6} \right)}.
\nonumber
\end{eqnarray}
This transformation is not exact and produces an overall systematic error 
proportional to $(\Delta \tau \lambda)^3$ in the Monte Carlo estimate of an observable. However, 
since we already have a systematic error proportional to $\Delta \tau ^2$ from the Trotter 
decomposition, the transformation
is as good as exact. It has the great advantage of being discrete thus
allowing efficient sampling.

\subsection{The Heisenberg limit.}
\label{Heisenberg}
We conclude this technical part with some comments concerning the numerical simulations of the 
Heisenberg model. At $t/J = 0$,  $H_U$  is a good quantum number since $[H_{tUJ},H_U] = 0$. Hence, 
in principle it suffices to choose a trial wave function $ | \Psi_T \rangle $   satisfying 
$H_U | \Psi_T  \rangle  = 0 $  to guarantee that the imaginary time propagation converges to 
the ground state of the Heisenberg model (see Eq. \ref{PQMC}). On the other hand, one can 
relax this constraint on the trial wave function  and 
implement a Gutzwiller projection  onto the Hilbert space with no double occupancy.  
We have found the second approach to be much more efficient.  

The algorithm we use here  is very related to the one we have used in Ref. \cite{Capponi00} where a
detailed technical section is provided.

\section{ Numerical results}
\label{Results}

\begin{figure}[h]
\begin{center}
\includegraphics[width=.4\textwidth]{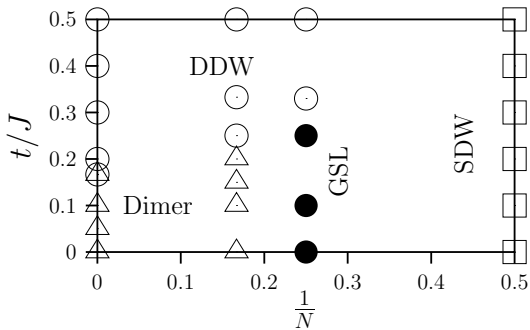} 
\end{center}
\caption[]{ Phase diagram of the half-filled 
(i.e. $\rho =  \frac{2}{N} \sum_{\alpha} \langle c^{\dagger}_{\vec{i},\alpha} c^{}_{\vec{i},\alpha} \rangle 
$) Hubbard-Heisenberg model as a function of $t/J$.  
For $t/J>0$ we set $U=0$. The $t=0$ line corresponds to the Heisenberg model where charge fluctuations 
are completely suppressed (see Sec. \ref{Heisenberg}). The symbols 
correspond to the parameters where we have carried out simulations and 
denote the following phases: $\bigtriangleup $: Spin-dimerized phase, $\bigcirc$: DDW phase, 
$\Box$: Spin-density wave phase, and $\bullet $: insulating phase with no broken lattice and 
spin symmetries and no gap to spin excitations (gapless spin-liquid (GSL)  phase). } 
\label{Phase.fig}
\end{figure}

Our results are summarized in the phase diagram shown in Fig. \ref{Phase.fig}. Here, we consider the 
half-filled case as a function of $N$ and $t/J$.  For values of $t/J > 0$ we  set $U=0$. The  $t=0$  
line corresponds to the Heisenberg model where 
charge fluctuations are completely suppressed (see Sec. \ref{Heisenberg}).
In the large-N  limit, the data stems for the mean-field calculation of the previous section.   At $N=6$, 
we essentially reproduce the saddle point result  with a somewhat smaller value of $t_c/J$ reflecting 
the instability of the  DDW phase in favor  of the spin-dimerized phase. 
Irrespective of the coupling $t/J$, the 
$SU(2)$ model shows an insulating spin-density wave (SDW) state. The most interesting  feature of the phase 
diagram occurs at $N=4$. Apart from the DDW phase present at {\it large} values of $t/J$ we find an 
insulating phase (solid circles in Fig. \ref{Phase.fig}) 
with no apparent broken symmetries and no spin gap.  We will argue that in this phase the antiferromagnetic 
spin correlations are critical  leading to gapless spin modes around the antiferromagnetic  wave vector 
$\vec{Q} = (\pi,\pi)$. Furthermore, we will present results showing that low lying  spin modes with 
very small  spectral weight  are present around the $\vec{q} = (0,\pi) $ and $\vec{q} = (0,\pi) $ 
wave vectors.  Before proceeding let us remind the reader that our simulations are carried out with
the projective algorithm of Eq. \ref{PQMC} and hence reflect ground state properties. 

To establish the above phase diagram, we have computed  equal-time and time displaced correlation 
functions.  Let $O(\vec{i}) $ be an observable, with time displaced correlation function: 
\begin{equation}
	S_O(\vec{i} - \vec{j}, \tau) = \langle O(\vec{i},\tau ) O ( \vec{j} ) \rangle 
                                   - \langle O(\vec{i})  \rangle \langle O ( \vec{j} ) \rangle 
\end{equation}   
and corresponding Fourier transform: 
\begin{equation}
	S_O(\vec{q}, \tau) = \sum_{\vec{r}} e^{i \vec{q}  \vec{r} } S_O(\vec{r},\tau). 
\end{equation}
From the equal time correlation function $ S_O(\vec{q}) \equiv S_O(\vec{q},\tau = 0) $, 
we can  establish the presence of long range order at a given wave vector. In this case 
$S_O(\vec{q})$ scales as the volume of the system, $V$, the proportionality constant being the 
square of the order-parameter.  
From the imaginary time displaced correlation functions, we can compute spectral functions, 
$S_O(\vec{q},\omega) $, by solving, 
\begin{eqnarray}
	S_O(\vec{q},\tau) & &  
        = \frac{1}{\pi} \int {\rm d} \omega S_O(\vec{q},\omega) e^{- \tau \omega } ,   \nonumber 
\\
	S_O(\vec{q}) & & = \frac{1}{\pi} \int {\rm d} \omega S_O(\vec{q},\omega),
\end{eqnarray} 
with the use of the Maximum Entropy method.

Information  on gaps and the  spectral weight of the lowest lying excitation,  is obtained 
directly form the imaginary time correlation function without having  recourse to  analytical 
continuation.  
\begin{eqnarray}
	S_O(\vec{q}, \tau) & = & \sum_n | \langle \Psi_0 | O(-\vec{q}) | \chi_n(\vec{q}) \rangle |^2 
e^{-\tau ( E_n(\vec{q}) - E_0 ) } \nonumber \\
	& \stackrel{\tau \rightarrow \infty}{\longrightarrow} & 
	 \underbrace{| \langle \Psi_0 | O(-\vec{q}) | \chi_0(\vec{q}) \rangle |^2}_
          {Z_O(\vec{q})} e^{ - \Delta_O(\vec{q}) \tau }.
\end{eqnarray}
In the above, $ | \Psi_0 \rangle $ corresponds to the normalized ground state.
$ | \chi_n(\vec{q}) \rangle $ are eigenstates of $H$  with momentum $\vec{q}$  and 
$| \langle \Psi_0 | O(-\vec{q}) | \chi_n(\vec{q}) \rangle | > 0$.  The  gap 
$\Delta_O(\vec{q}) $   corresponds to the energy difference between the first excited state 
$| \chi_0(\vec{q}) \rangle $  and the ground state and    
$ O(\vec{q}) = \frac{1}{\sqrt{V}} \sum_{\vec{j}} e^{i \vec{q} \vec{j} } O(\vec{j}) $. Finally, 
the  residue $Z_O(\vec{q})$  corresponds to the spectral weight of the lowest lying excitation.

\begin{figure}
\begin{center}
\includegraphics[width=.45\textwidth]{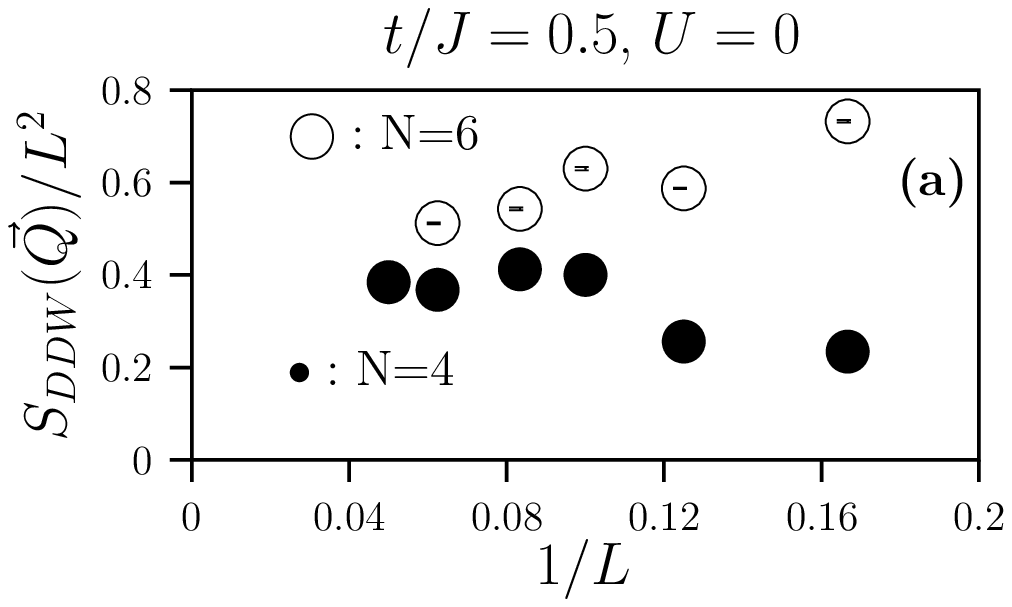} \\ 
\vspace{0.2cm}
\includegraphics[width=.3\textwidth, height=0.29\textheight]{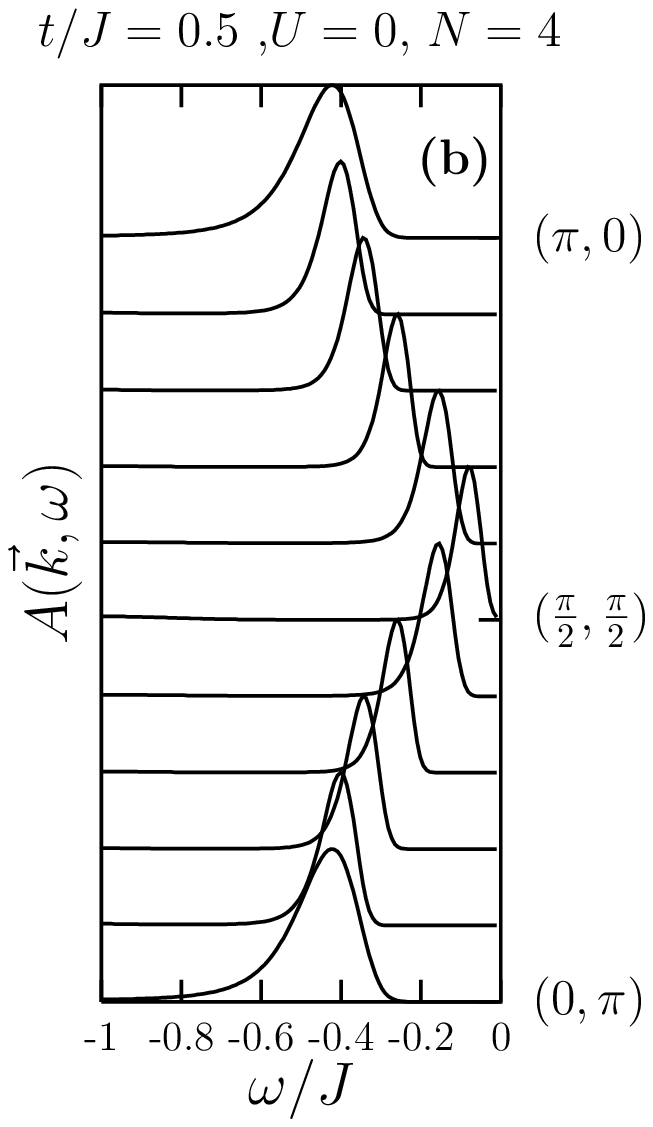} \\
\vspace{0.2cm}
\includegraphics[width=.45\textwidth]{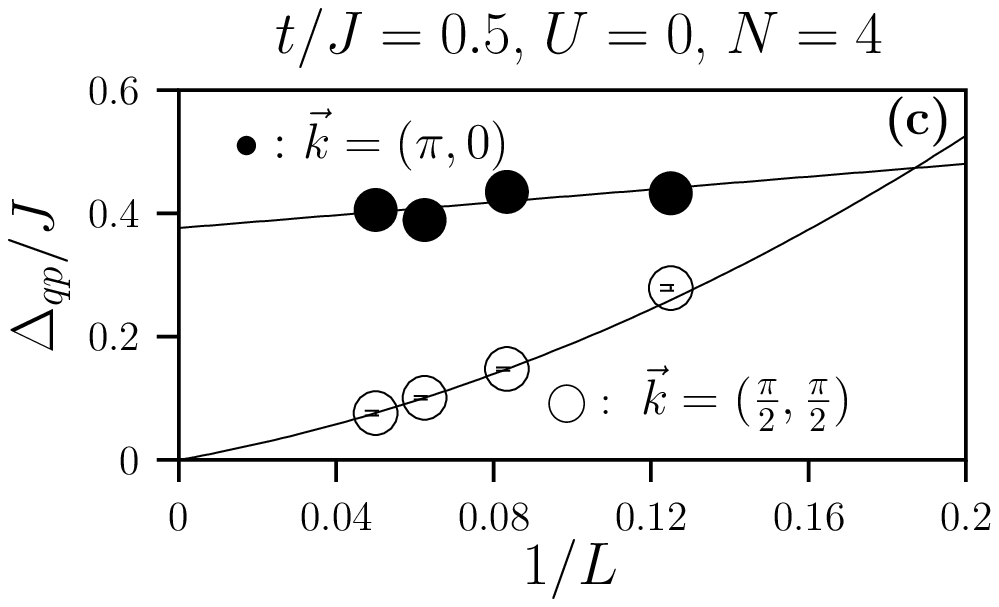} 
\end{center}
\caption[]{ (a) Size scaling of the DDW equal time correlation functions at wave vector 
$\vec{Q} = (\pi,\pi)$. (b)  Single particle spectral function in the DDW phase on a $20 \times 20$ lattice. 
Here, we normailze the peak height to unity and along the $(0,\pi)$ to $(\pi,0)$ line 
each spectrum satisfies the sum-rule $\int_{-\infty}^{0} {\rm d} \omega A(\vec{k},\omega)  = \pi n(\vec{k}) = \pi/2$.
(c) Size scaling
of the quasiparticle gap at $\vec{k} = (\pi/2,\pi/2) $ and $\vec{k} = (0,\pi)$. The data 
shows the semi-metal character of the DDW phase.   }
\label{DDW.fig}
\end{figure}

\begin{figure}
\begin{center}
\includegraphics[width=.45\textwidth]{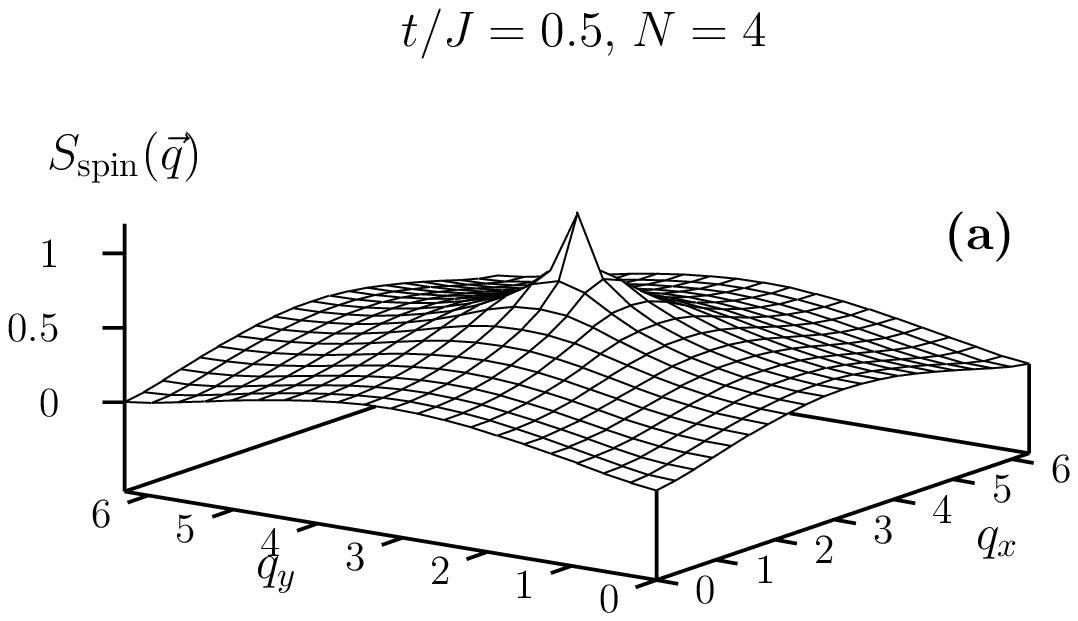} \\
\vspace{0.2cm}
\includegraphics[width=.45\textwidth]{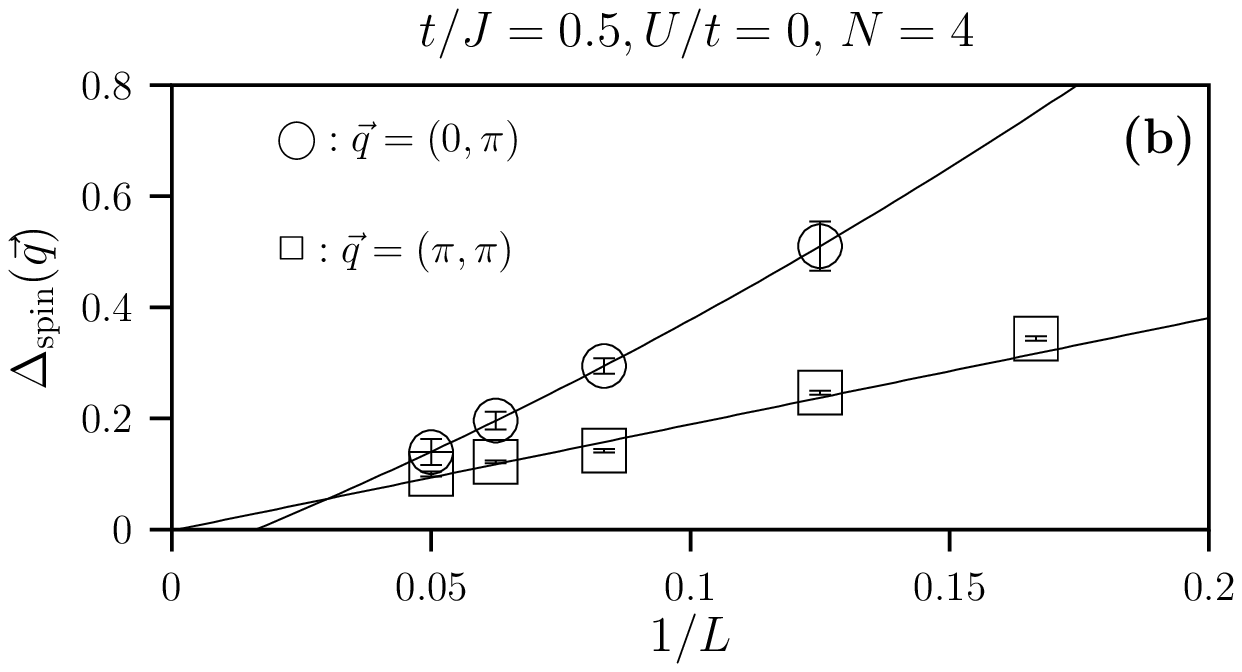} 
\end{center}
\caption[]{Spin correlations in the DDW phase. a) Equal time spin structure factor. b)  Size scaling of the 
spin gap at $\vec{q} = (0,\pi) $ and $\vec{q} = (\pi,\pi)$. }
\label{DDW_spin.fig}
\end{figure}

To study the model, we have considered the following observables. 
Let us define the magnetization as
\begin{equation}
	O_{\rm spin} (\vec{i}) =  \sum_{\alpha} f(\alpha) c^{\dagger}_{\alpha} c_{\alpha} \; \; {\rm with} 
\; \; \sum_{\alpha} f(\alpha) = 0. 
\end{equation}
For even values of $N$ considered here, we choose $f(\alpha) = \pm 1$. 
Note that  $SU(N)$ symmetry leads to the identity  
\begin{equation}
   S_{\rm spin}(\vec{i} - \vec{j} )  \equiv \langle 	O_{\rm spin} (\vec{i})  O_{\rm spin} (\vec{j}) 
\rangle = \frac{N}{N^2 -1} \langle \sum_{\alpha,\beta} S_{\alpha,\beta,\vec{i}}  S_{\beta,\alpha,\vec{j}} 
\rangle
\end{equation}
where $ S^{\alpha,\beta}_{\vec{i}} $ are the generators of $SU(N)$ (See Eq.  \ref{generators} ).

To detect spin-dimerization and DDW instabilities we consider respectively
\begin{equation}
	O_{\rm dimer}(\vec{i}) = O_{\rm spin}(\vec{i}) O_{\rm spin}(\vec{i} + \vec{a}_x ) 
\end{equation}
and 
\begin{equation}
	O_{\rm DDW}(\vec{i}) =   j_x(\vec{i}) - j_y(\vec{i}) 
\end{equation}
with current: 
\begin{equation}
	j_x(\vec{i}) = i \sum_{\alpha} 
   \left( c^{\dagger}_{\vec{i},\alpha} c_{\vec{i} + \vec{a}_x,\alpha}  - 
          c^{\dagger}_{\vec{i} + \vec{a}_x,\alpha} c_{\vec{i} ,\alpha}   \right)
\end{equation}
and an equivalent form for $j_y(\vec{i})$.
Finally, we obtain information on single particle excitations by measuring 
time displaced Green functions: 
$G(\vec{k},\tau) = -\langle T c_{\vec{k},\alpha }(\tau) c^{\dagger}_{\vec{k},\alpha} (0)  \rangle $.
From this quantity we can extract quasiparticle gaps   as well as the spectral function 
$A(\vec{k}, \omega)$.

\subsection {The DDW phase}
We start our description of the phase diagram with the DDW phase. Fig. \ref{DDW.fig}a shows the 
finite size scaling of $S_{\rm DDW} ( \vec{Q} ) /L^2 $  at the antiferromagnetic wave vector 
$\vec{Q} = (\pi,\pi) $.  For $t/J = 0.5$ and both considered values of $N$  the data supports 
$\lim_{L \rightarrow \infty} S_{\rm DDW} ( \vec{Q} ) /L^2 > 0$ thus signaling a DDW ordered phase. 
One can confirm this point of view from the analysis on the single particle spectral function 
along the $ \vec{k} = (0,\pi) $ to $ \vec{k} = ( \pi, 0) $ line in the Brillouin zone  (see 
Fig. \ref{DDW.fig}b). After size scaling (see Fig. \ref{DDW.fig}c) the data is consistent with 
the vanishing of the quasiparticle gap at $ \vec{k} = ( \pi/2,\pi/2) $ thus confirming the 
semi-metal character of the DDW phase. 

Since the single particle spectrum has gapless excitations at the nodal points  
$\vec{k} = ( \pm \pi/2, \pm \pi/2) $ we can  expect gapless spin excitations  centered around the
wave vectors $\vec{q} = (0,\pi)$, $\vec{q} = (\pi,0) $ and $\vec{q} = (\pi,\pi)$, 
along with a power-law decay of the equal time spin-spin correlations. 
Fig. \ref{DDW_spin.fig} confirms this. The spin gap vanishes at 
the above mentioned  wave vectors (Fig. \ref{DDW_spin.fig}b). 
The spectral weight, $Z_{\rm spin} (\vec{q})$,  of the low lying $ \vec{q} 
= (0,\pi) $  spin excitation  is very small in comparison to $ \vec{q} = (\pi,\pi)$. For the 
$L=20$ lattice we have  $Z_{\rm spin}(\vec{q})/S_{\rm spin}(\vec{q}) \simeq 0.006 $  at 
$ \vec{q}  = (0,\pi) $ and 
$Z_{\rm spin} (\vec{q})/S_{\rm spin} (\vec{q}) \simeq 0.5 $  
at $ \vec{q}  = (\pi,\pi) $.  The data on which this statement is based stems  from  Fig. 
\ref{Sqom_N4.fig}c.
The size scaling  of the static spin structure factor at $\vec{Q} = (\pi,\pi)$
is consistent with a power-law decay of the staggered spin-spin correlation functions: 
$S_{\rm spin}( \vec{Q}= (\pi,\pi) ) /L^2 \propto L^{-2} $. 
It is interesting to note that even though gapless spin excitations are present at  $\vec{q} = (0,\pi)$ 
no sign of a cusp in the spin structure factor at this wave vector is apparent 
(See Fig. \ref{DDW_spin.fig}a).  This signals a  {\it large}  exponent for the $(0,\pi)$ spin correlations. 
Note that in the saddle approximation both the $(0,\pi)$ and $(\pi,\pi)$ spin correlations decay as 
$r^{-4}$ and no cusp feature in the static spin structure factor is apparent \cite{Arovas88}.

\subsection{N=6} 

\begin{figure}
\begin{center}
\includegraphics[width=.45\textwidth]{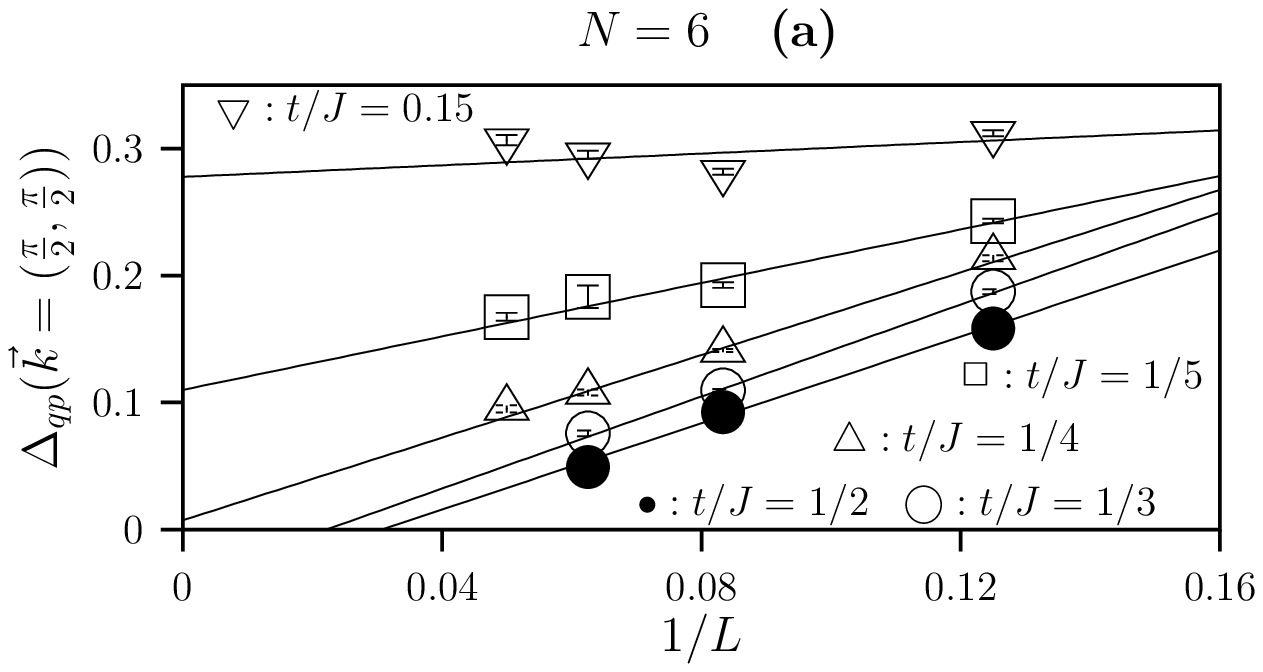} \\ 
\vspace{0.2cm}
\includegraphics[width=.45\textwidth]{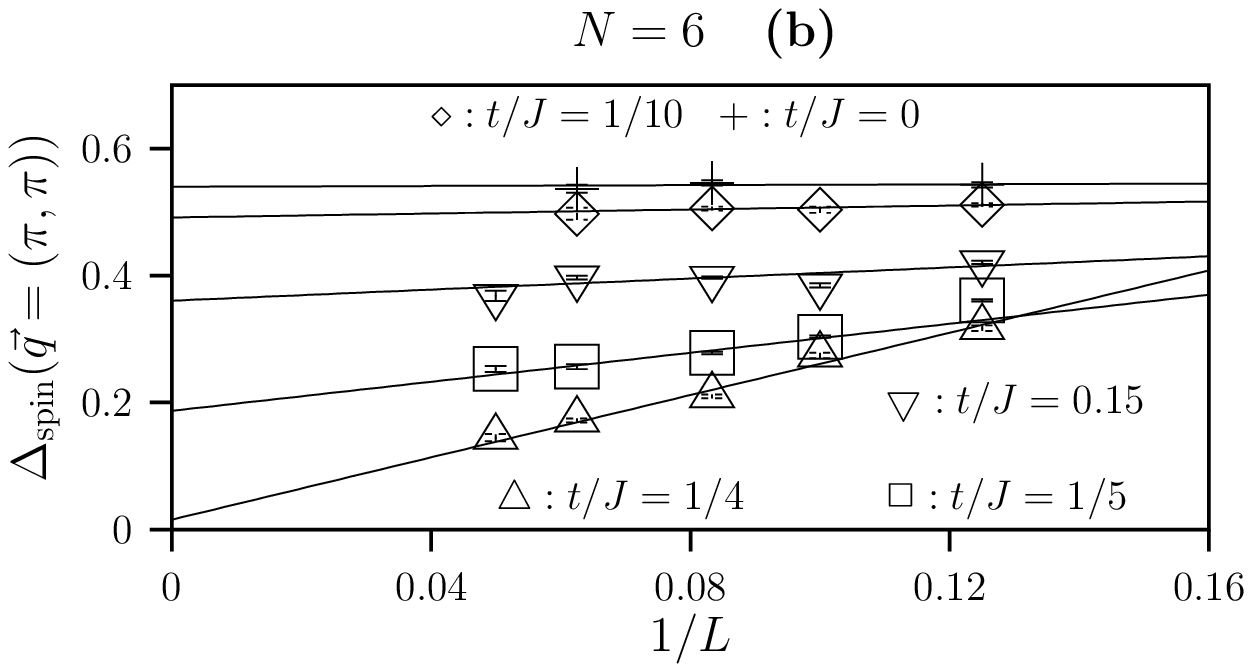} 
\end{center}
\caption[]{Quasiparticle (a)  and spin (b) gaps at $N=6$ as a function of coupling. 
For $t/J > 0$ we set $U=0$, and 
$t/J =0$ refers to the Heisenberg model (see Sec. \ref{Heisenberg}). }
\label{Gaps_N6.fig}
\end{figure} 

To study the stability of the  DDW phase from $t/J = 0.5$ to the Heisenberg point,  
it is convenient to consider the quasiparticle gap at $ \vec{k} = \left( \pi/2, \pi/2 \right) $ 
(see Fig. \ref{Gaps_N6.fig}a). For values of $t/J \leq 1/5$ the data   is consistent 
with the opening of a quasiparticle gap at  $\vec{k} = (\pi/2,\pi/2)$.   
The opening of the quasiparticle gap is accompanied  by the opening of a spin-gap (see \ref{Gaps_N6.fig}b).
Hence, the data suggest that for values of $t/J \leq 1/5$ at $N=6$ we have entered a spin dimerized phase, 
with spin and charge gaps.

\begin{figure}
\begin{center}
\includegraphics[width=.4\textwidth]{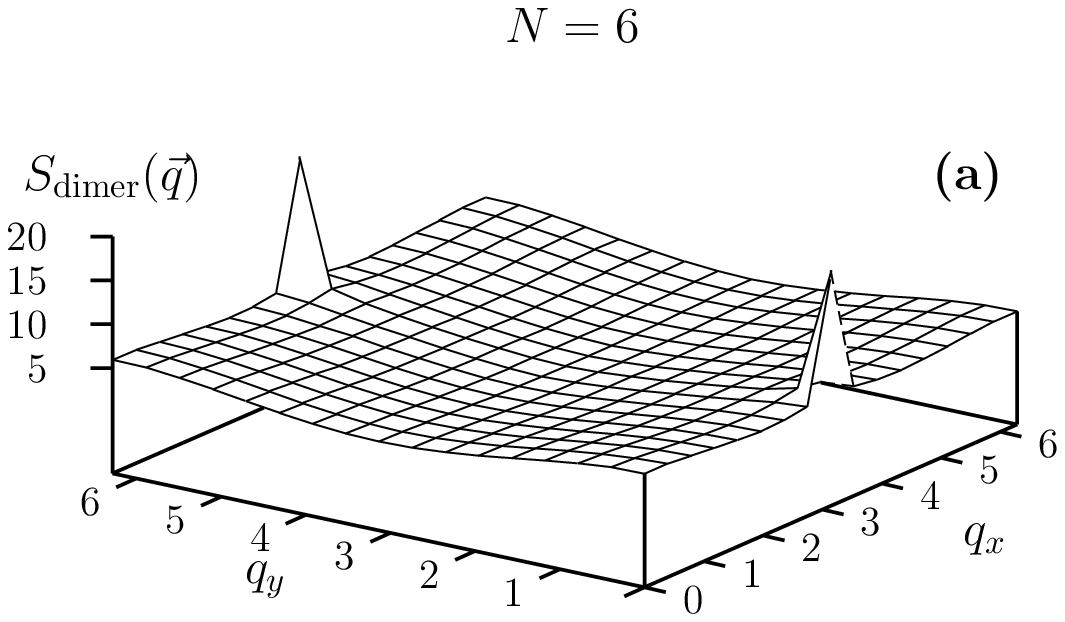} \\ 
\vspace{0.2cm}
\includegraphics[width=.45\textwidth]{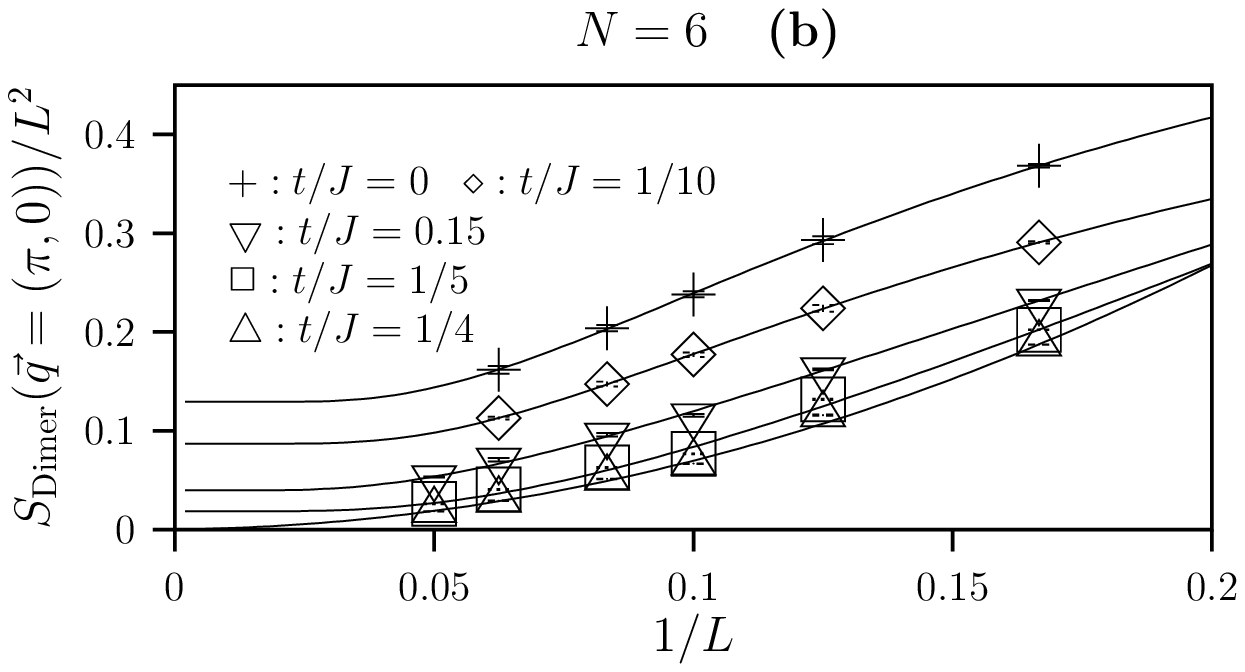} \\ 
\end{center}
\caption[]{ (a) Equal time dimer correlation functions for the $SU(6)$  Heisenberg model. (b) Size 
scaling of the dimer correlation function at $\vec{q} = (0,\pi)$. For $t/J > 0$ we  set $U=0$, and 
$t/J =0$ refers to the Heisenberg model (see Sec. \ref{Heisenberg}). }
\label{Corr_N6.fig}
\end{figure}

We confirm this point of view by  looking into the dimer correlation functions 
(see Fig. \ref{Corr_N6.fig}a). Those correlations are dominant at wave vector 
$\vec{q} = (\pi,0) $ in agreement with the mean-field dimerization pattern. To establish 
long-range order  we have to extrapolate $S_{\rm dimer} (\pi,0)/L^2 $ to the thermodynamic limit. 
For $t/J \leq 1/5$  gaps are present both in the charge and spin degrees of freedom.
Since the presence of
gaps is equivalent to the localization of spin and charge degrees of freedom, we fit the 
finite size data  to the form: $ a  +  br^c e^{-L/\xi} $.   Adopting this form for the finite size 
corrections, the data is consistent with the onset of a spin-dimerized state  for $t/J \leq 1/5$.

\subsection{N=4} 

\begin{figure}
\begin{center}
\includegraphics[width=.45\textwidth]{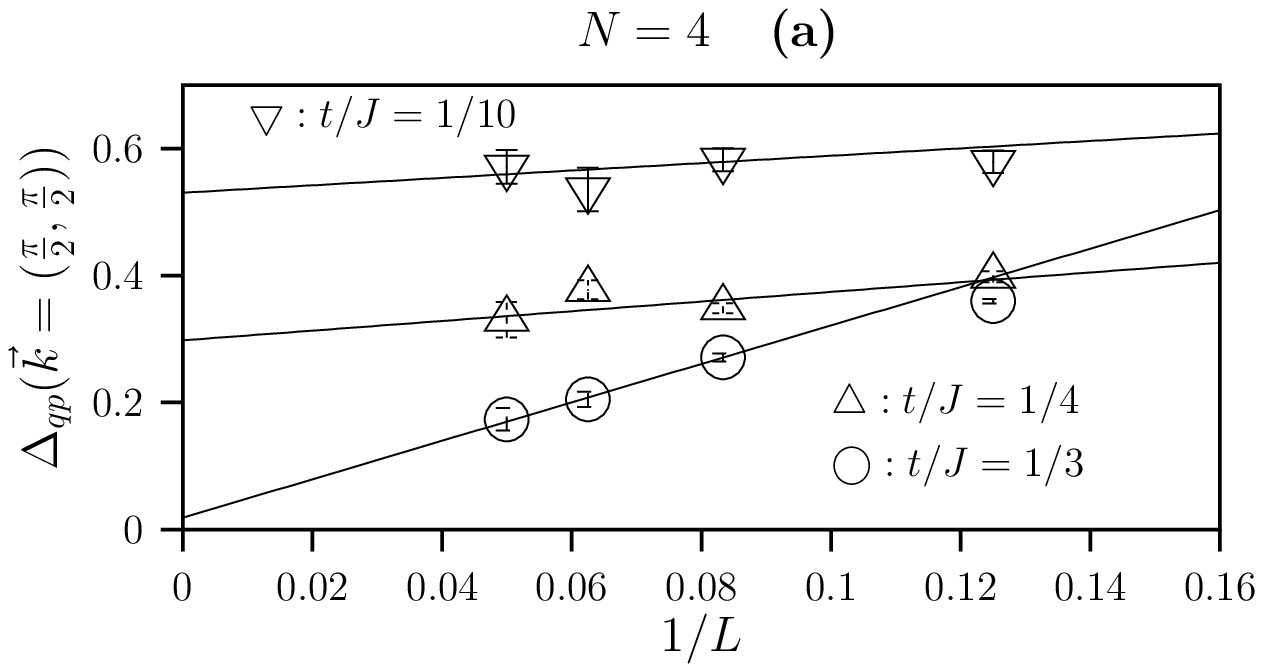} \\ 
\vspace{0.2cm}
\includegraphics[width=.45\textwidth]{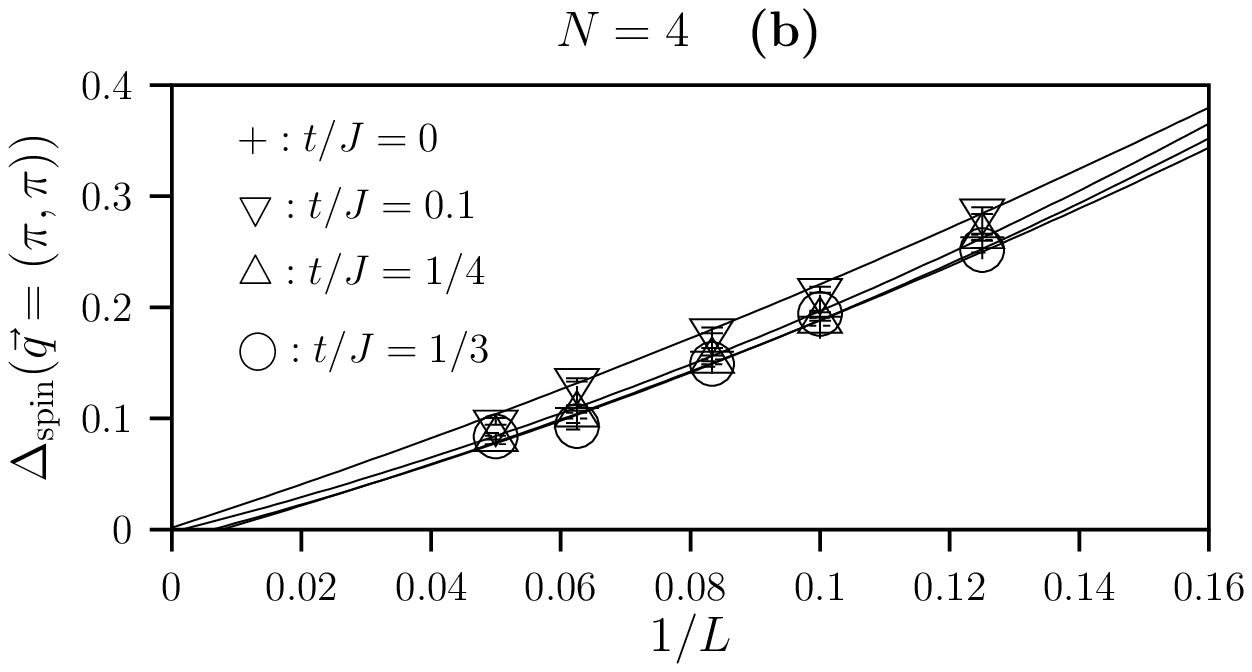} 
\end{center}
\caption[]{Size scaling of the quasiparticle and spin gaps at $N=4$. 
For $t/J > 0$ we set $U=0$, and 
$t/J =0$ refers to the Heisenberg model (see Sec. \ref{Heisenberg}).}
\label{Gaps_N4.fig}
\end{figure}

\begin{figure}
\begin{center}
\includegraphics[width=.4\textwidth]{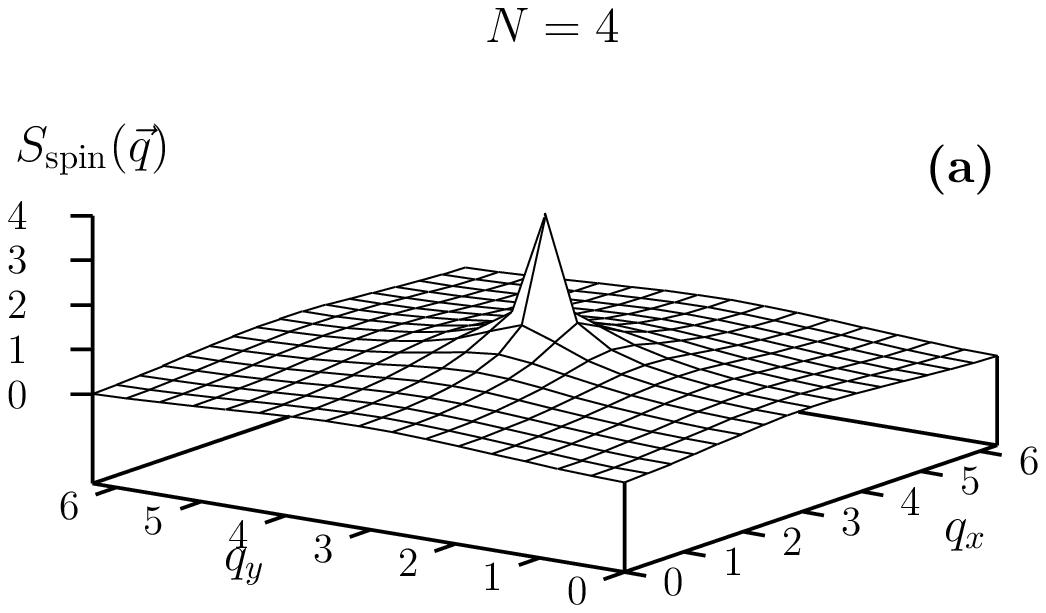} \\ 
\includegraphics[width=.4\textwidth]{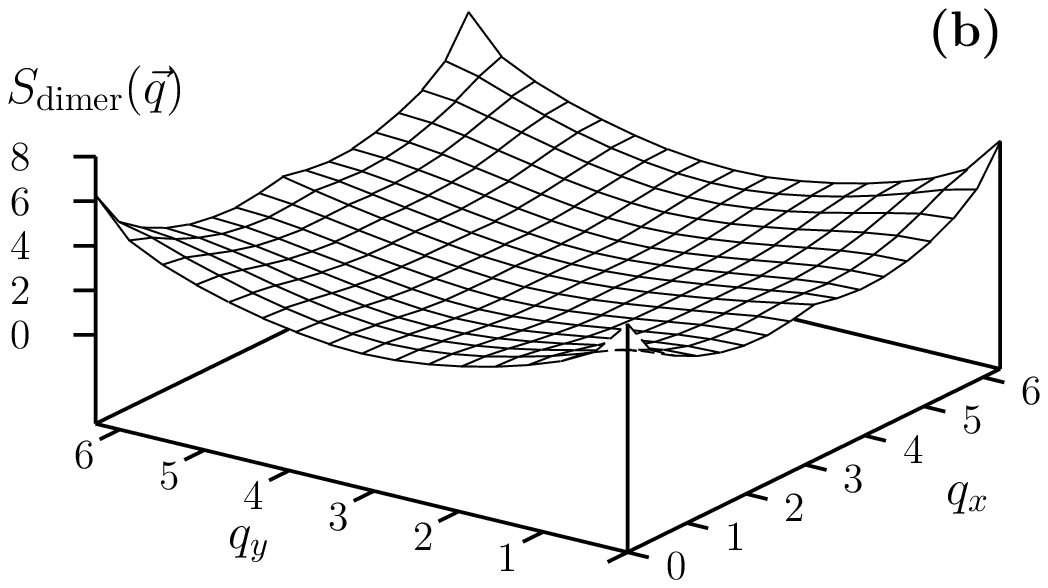} 
\end{center}
\caption[]{ Equal time dimer and spin correlation functions for the $SU(4)$ Heisenberg model. }
\label{N4_cor.fig}
\end{figure}

We now turn our attention to the $SU(4)$ model.  From the size analysis of the quasiparticle gap at 
$\vec{k} = (\pi/2,\pi/2) $ (see Fig. \ref{Gaps_N4.fig}a) we can conclude that the DDW state is 
unstable  for values of $t/J \leq 1/4$. In contrast to the $N=6$ case, no spin gap opens across 
the transition up to the Heisenberg limit (see Fig. \ref{Gaps_N4.fig}b).  
The correlation functions presented in Fig. 
\ref{N4_cor.fig} in the Heisenberg limit show dominant spin-spin correlations. In accordance with 
the absence of spin gap no sign of long-range dimerization is apparent. 

\begin{figure}
\begin{center}
\includegraphics[width=.45\textwidth]{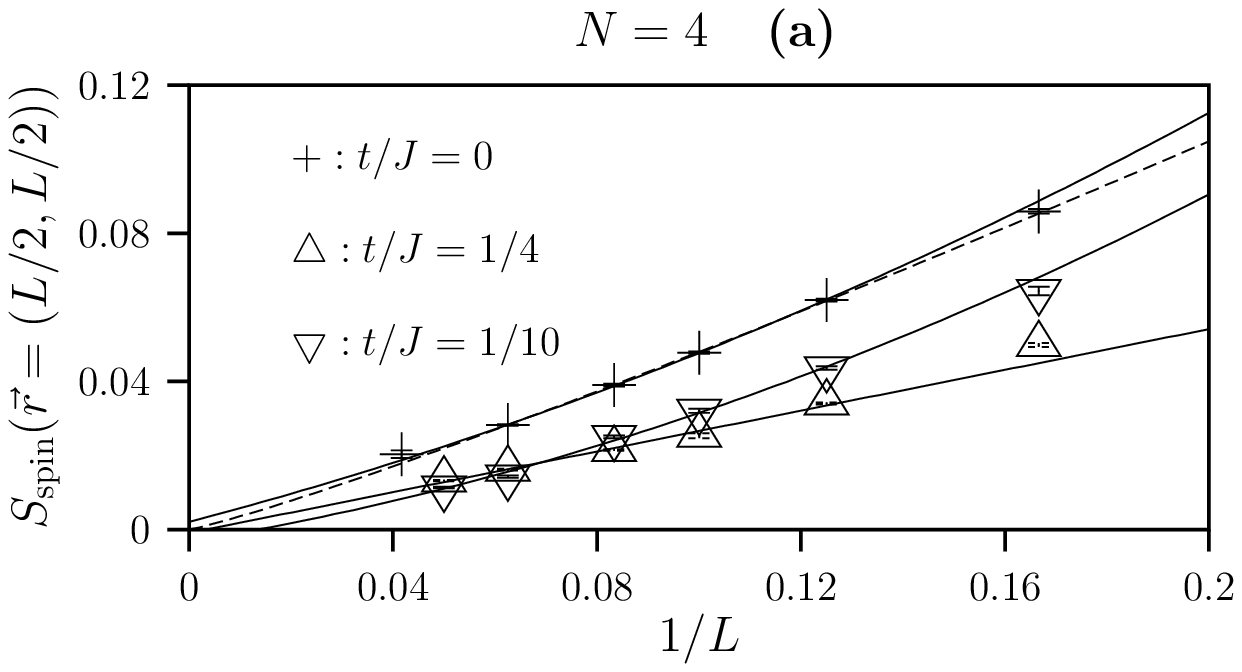} \\ 
\includegraphics[width=.45\textwidth]{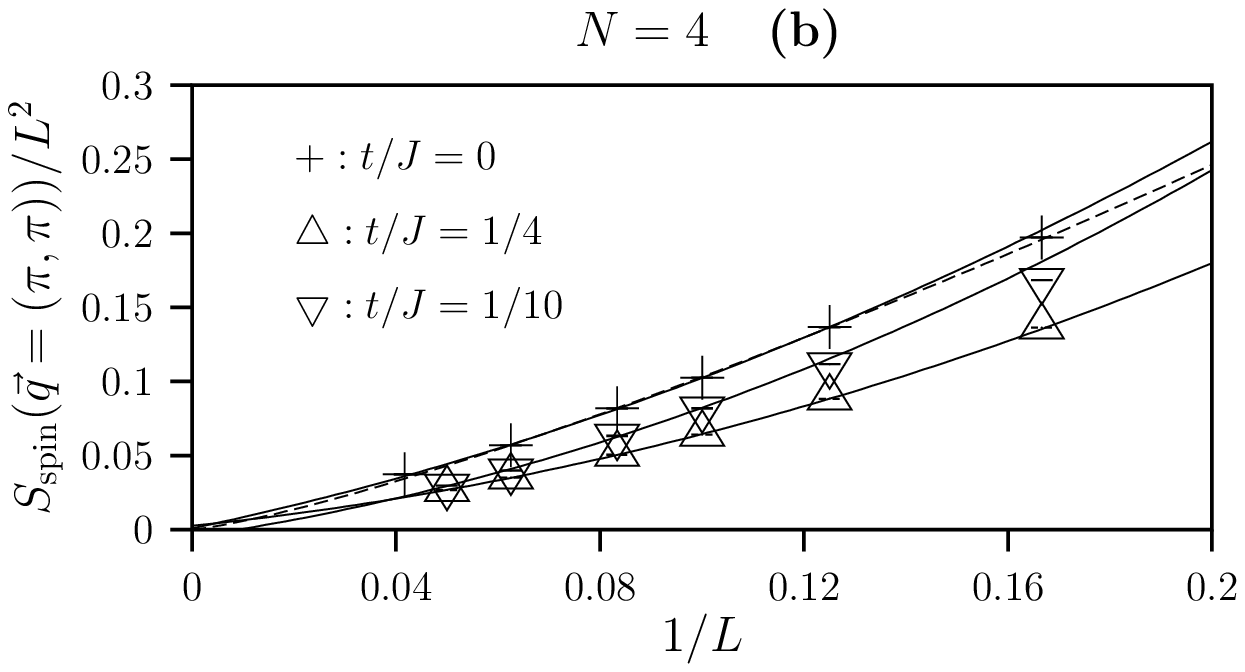} 
\end{center}
\caption[]{ Size scaling of the equal time spin-spin correlation functions.
For $t/J > 0$ we set $U=0$, and 
$t/J =0$ refers to the Heisenberg model (see Sec. \ref{Heisenberg}).}
\label{Spin_N4.fig}
\end{figure}

The issue is now to establish the existence or non-existence of long range antiferromagnetic 
order. In Fig. \ref{Spin_N4.fig}, we plot the spin-spin correlation at the largest distance, 
$(L/2,L/2)$, on our $L \times L $ lattice  as well as 
$S_{\rm spin}(\vec{q} = ( \pi,\pi) )/L^2 $ as a function of $1/L$. As apparent, the extrapolation 
is consistent with the absence of long-range magnetic ordering. In particular, at the Heisenberg point, 
where we have carried out simulations on lattices ranging up to $24 \times 24 $, our extrapolated values 
read $\lim_{L \rightarrow \infty} S_{\rm spin}  ( \vec{r} = \left( L/2,L/2\right) ) = 0.002 \pm 0.003$ 
along with 
$ \lim_{L \rightarrow \infty }  S_{\rm spin}(\vec{Q} = (\pi,\pi) )/ L^2   = 0.0008 \pm  0.004 $. 

The data is equally consistent with a power-law decay of the spin-spin correlations. Concentrating again 
on the Heisenberg point the dashed lines in Fig. \ref{Spin_N4.fig} correspond to the forms: 
$ S_{\rm spin}(\vec{q} = ( \pi,\pi) )/L^2  \propto L^{-1.25} $ and 
$ S_{\rm spin}( \vec{r} = (L/2,L/2)  )  \propto L^{-1.12} $. The difference between the two numerical 
values of the  exponents gives an idea of their  uncertainty.  

\begin{figure}
\begin{center}
\includegraphics[width=.35\textwidth]{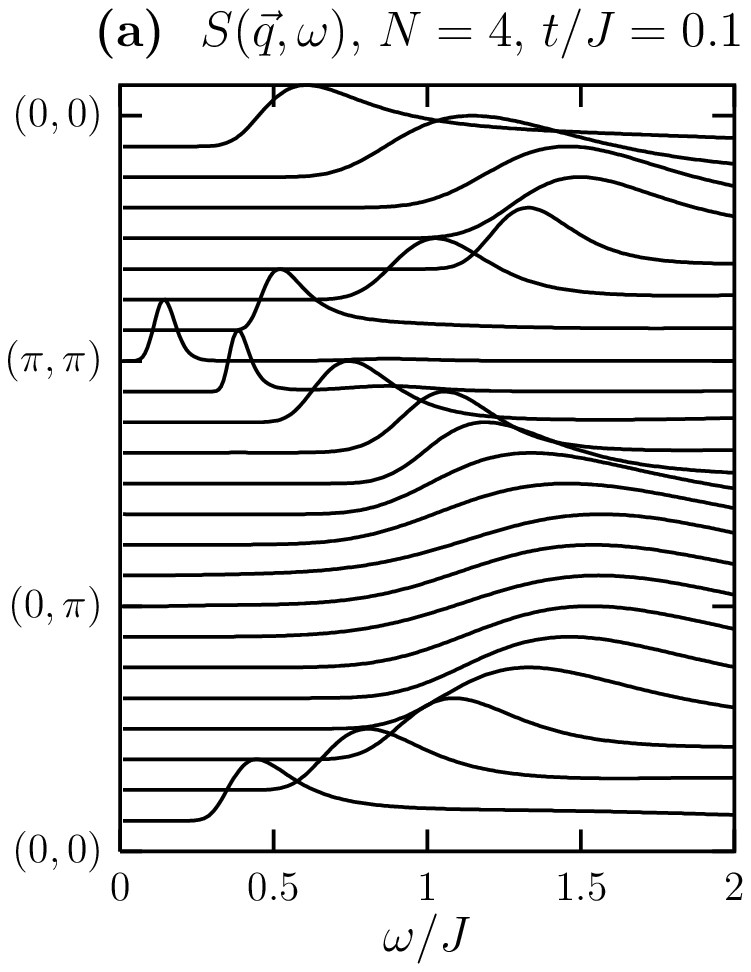} \\ 
\vspace{0.1cm}
\includegraphics[width=.49\textwidth]{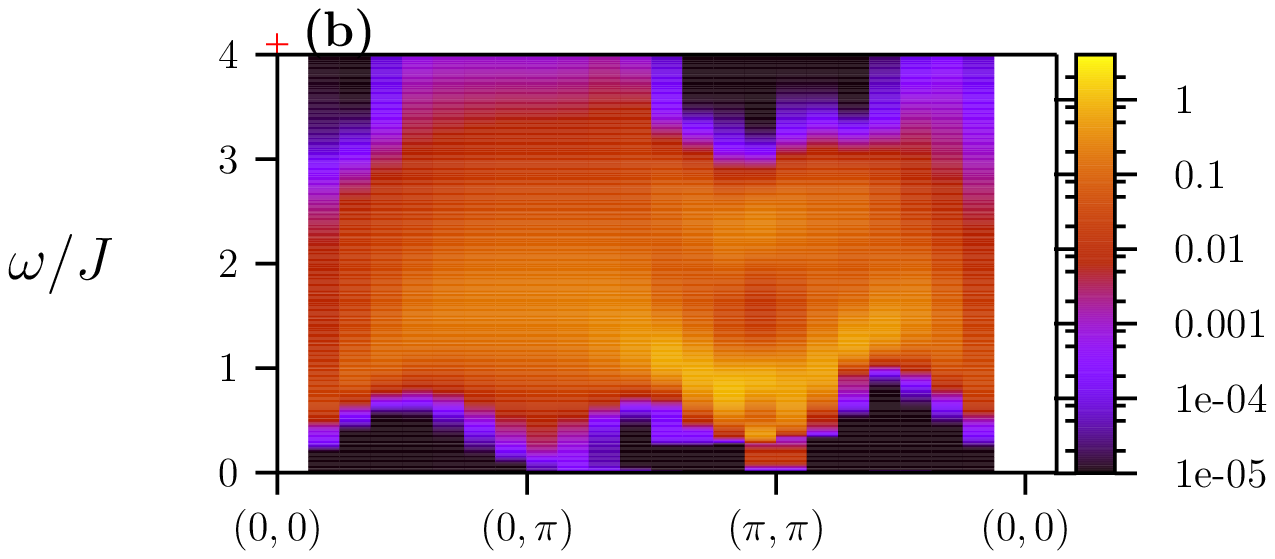} \\ 
\vspace{0.2cm}
\includegraphics[width=.47\textwidth]{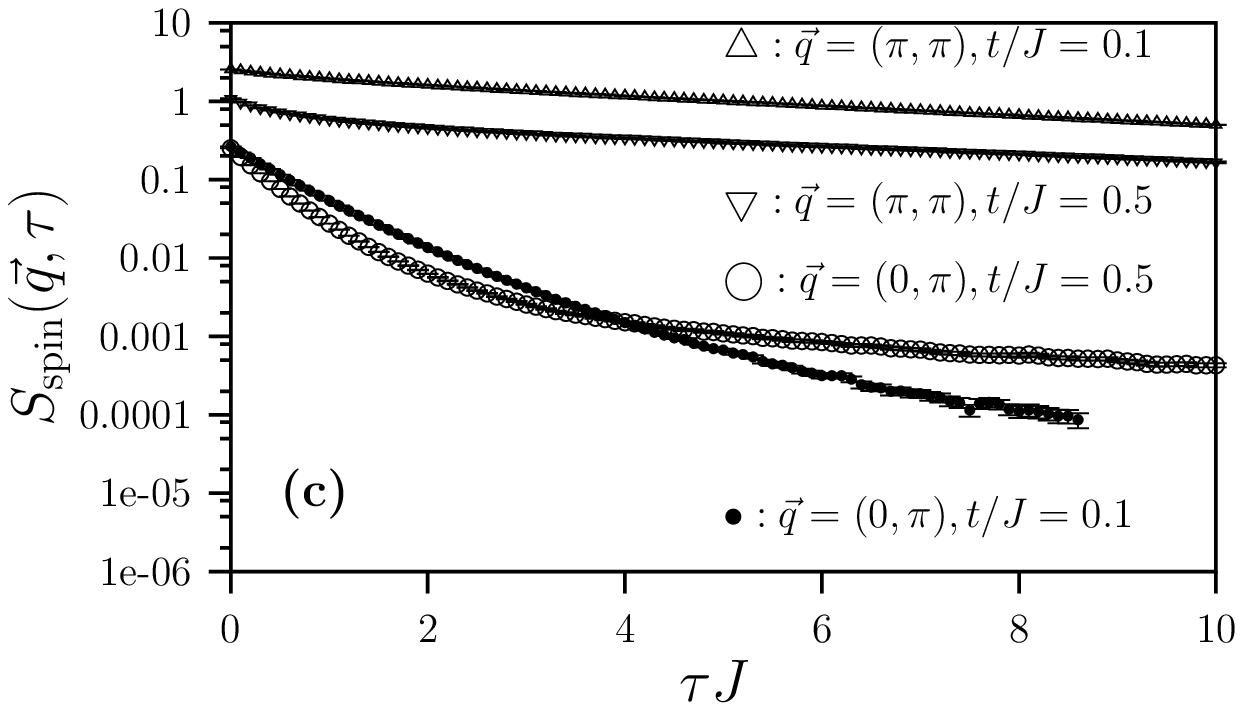} 
\end{center}
\caption[]{ (a) Dynamical spin structure factor on a $16 \times 16$ lattice. 
We have normalized the peak height of each spectrum 
to unity so as to put forward the overall shape of the dispersion relation. The reader however has to 
bear in mind that the   weight under each spectrum is equal to the equal structure factor 
$S_{\rm spin}(q)$ which is peaked at $\vec{q} = (\pi,\pi) $ and vanishes at $\vec{q} = (0,0)$. 
(b) Intensity plot of the data of (a)  on a logarithmic scale. 
(c) Imaginary time correlation functions.  Here, one can see that without having recourse to 
analytical continuation,  the  bare imaginary time data supports the existence of low lying modes 
at $\vec{q} = (0,\pi)$ (See text).  At $t/J = 0.5$ ($t/J= 0.1$) we consider a $20 \times 20$ 
($16 \times 16$) lattice. }
\label{Sqom_N4.fig}
\end{figure}

Hence the equal time data is consistent with an insulating phase with no apparent lattice 
or spin symmetry breaking and no gap in the magnetic excitations.
It is now intriguing to  investigate the spin-dynamics of this phase.  Fig. \ref{Sqom_N4.fig}a
plots the dynamical spin-structure factor at $t/J = 0.1$, $U/t = 0$ on a $16 \times 16 $ lattice.  
The data, shows several features.  The gap at  $\vec{q} = (\pi,\pi)$ is a finite size effect  
(see Fig. \ref{Gaps_N4.fig}b). Taking this into account, the data is consistent with  a gapless 
mode with linear dispersion around $\vec{q} = (\pi,\pi)$.  This feature is clearly not surprising since 
the equal time correlation functions show critical behavior at this wave vector.   
As we follow this mode to $\vec{q} = (0,\pi)$  the 
line shape becomes very broad and  spectral weight seems to spill down 
to low energies.  This is especially apparent on the intensity plot for which we have used a
logarithmic scale (see Fig. \ref{Sqom_N4.fig}b).
Since the spectral weight of the low-lying modes around $\vec{q} = (0,\pi)$ is very small, it is 
desirable  to confirm   the above statement.  To this aim we plot in Fig. \ref{Sqom_N4.fig}c   the 
imaginary time  displaced correlation functions, 
$S_{\rm spin} (\vec{q},\tau)$,  at $\vec{q} = (0,\pi) $ and $\vec{q} = (\pi,\pi)$ 
both in the gapless spin-liquid phase at $t/J = 0.1$ and for comparison in the DDW phase at $t/J = 0.5$.
Let us start with the  DDW phase  where we were able to show the presence of gapless  spin 
modes  around the $(\pi,\pi)$, $(0,\pi) $ and $(\pi,0) $ points in thermodynamic limit 
(see Fig. \ref{DDW_spin.fig}b). On the $L=20$ sized system considered in Fig. \ref{Sqom_N4.fig}c one sees
that both the  $\vec{q} = (0,\pi)$  and 
$\vec{q} = (\pi,\pi)$ correlators decay assymtotically with the same exponential form, thus signalling 
low energy spin excitations at both wave vectors.  Of course there is a big difference in the prefactor
mutiplying this exponential decay. This replects the fact that the spectral weight of the low-lying 
$(0,\pi)$ excitation is much smaller than that of the $\vec{q} = (\pi,\pi)$ excitation.  
Let us now turn our attention to the gapless spin liquid phase at 
$t/J = 0.1$. As apparent the data shows a very similar behavior at the exception that the 
spectral weight at $\vec{q} = (0,\pi)$ is reduced  in comparison to the DDW phase. We note that due to the extremely 
small scales involved in the $\vec{q} = (0,\pi) $ data at $t/J = 0.1$ we are unable to obtain accurate 
results beyond $\tau J  \simeq 8 $.
Hence, on the basis of this 
data, we believe that the spin-liquid phase indeed shows low lying spin  modes not only at  $(\pi,\pi)$, 
but also at $(0,\pi)$ and $(\pi,0)$.  Finally, we note that as we approach the Heisenberg point, 
the spectral weight of the low lying mode at $\vec{q} = (0,\pi)$ becomes smaller and smaller.  This 
renders numerical detection of this feature at the Heisenberg point hard.  

\section {Summary and discussion.}

The half-filled  $SU(N)$ Hubbard Heisenberg model  shows a variety of phases, which we have 
analyzed in detail.  The saddle point physics --  a DDW phase at large values of  $t/J$ which 
becomes unstable to a spin dimerized  state  --  is valid down to $N=6$.  On the other hand, the 
$SU(2)$ model has a SDW insulating phase irrespective of the coupling constant.  
The most intriguing aspect of the phase diagram, is the gapless spin liquid state in the $SU(4)$ 
model in the vicinity of the Heisenberg point.  
The $SU(4) $ model has a DDW ground state at large values of $t/J$.  As appropriate for this 
semi-metallic state, we find gapless single particle excitations at wave  vectors 
$ \vec{k} = ( \pm \pi/2, \pm \pi/2 ) $. In the particle-hole channel those single particle 
excitations lead to gapless spin modes  centered around $ \vec{q} = (\pi,\pi) $,   $\vec{q} = (0,\pi) $, 
and  $ \vec{q} = (\pi,0) $. Reducing the magnitude of the hopping 
matrix element we find a semi-metal to insulator transition. In the insulating phase the 
antiferromagnetic, $\vec{Q} = (\pi,\pi)$,  spin correlations are critical and for the $SU(4)$ 
Heisenberg  model the data is consistent with the form 
$S_{\rm spin } (\vec{r} ) \sim e^{i \vec{Q} \cdot \vec{r} } |\vec{r} |^{-1.12}$.  This state shows 
no lattice broken symmetries and hence is a candidate for a gapless spin liquid state. 
In the particle-hole channel, the dynamical spin structure factor points  to gapless excitations  at 
$\vec{q} = ( \pi,\pi)$  but also to  low-lying modes with small spectral weight centered around  
$\vec{q} = (0,\pi) $ and  $\vec{q} = (\pi,0) $. 

It is  tempting to argue  that the gapless spin liquid phase   is well described by a  
DDW mean-field state 
supplemented with a Gutzwiller projection.  Requiring  invariance under time 
reversal symmetry pins the flux in each elementary square to $\pi$.   
Clearly, the  Gutzwiller projection triggers the semi-metal to insulator transition but one could argue 
that in  the particle-hole channel excitations remain gapless such that low lying 
spin excitations are present at wave vectors $\vec{q} = (\pi,\pi)$, $\vec{q} = (0,\pi)$, and $q=(\pi,0)$.
In the $SU(2) $ case, this variational wave function has been investigated in details. 
It turns out that due to a local  $SU(2)$ symmetry  \cite{Affleck88a} it  is equivalent to a 
BCS $d-$wave Gutzwiller projected wave 
function \cite{Gross88}. At the particle-hole symmetric point the equal time spin-spin correlations 
of this wave function have  been computed  \cite{Paramekanti04} to obtain a power-law decay 
of the antiferromagnetic correlations: $S(\vec{r}) \sim e^{i \vec{Q} \cdot \vec{r} } |\vec{r}|^{-1.5} $. 
Furthermore,  the spin structure factor as computed form the variational wave function shows no 
cusp feature at $\vec{q} = (0,\pi)$ and $\vec{q} = (\pi,0)$  
\footnote{A. Paramekanti, private communication}. 
This result compares favorably with 
behavior of the spin-spin correlations in the $SU(4) $ Heisenberg model discussed in the present work.

Alternatively we can ask the question of whether or not our results for the $SU(4)$ Heisenberg model 
are understandable from the perspective of the large-N saddle point. 
In the large-$N$ limit and at the  Heisenberg point one can stabilize the  $\pi$-flux  phase 
by adding biquadratic terms to the Hamiltonian \cite{Marston89}.
Furthermore, Hemerle et al. \cite{Hermele04}  have argued in favor of the stability of 
the $\pi$-flux phase to gauge fluctuations arising from the constraint of no double occupancy. 
At the mean-field level,  the $\pi$-flux phase shows antiferromagnetic spin-spin correlations 
which decay as  $S(\vec{r})  \sim e^{i \vec{Q} \cdot \vec{r} }  |\vec{r}|^{-4}$. 
Including gauge fluctuations in a $1/N$ approximation reduces the mean field exponent \cite{Rantner02}.
On the other hand, the $ (0,\pi) $ spin-spin correlations remain unaffected by gauge fluctuations 
\cite{Rantner02}.   Those results compare favorably with our  calculations at $N=4$.

To  investigate  the nature of the spinless liquid state we find in the $SU(4) $ Heisenberg model, 
it is desirable to investigate it's  behavior under  perturbations.  Following the variational work 
of \cite{Paramekanti04,Ivanov02}, the particle-hole symmetric point is unstable towards a $Z_2 $ spin 
liquid as realized for example in quantum dimer models \cite{Misguich02}. 
This instability is triggered by the inclusion of 
a next nearest neighbor hopping matrix element in the BCS Slater determinant. In the  spin 
model this translates in 
the inclusion of a  frustrating  exchange  coupling. Unfortunately, this  is not accessible 
to the quantum Monte Carlo approach since frustration leads to a minus sign problem. Another perturbation, 
which is accessible to the Monte Carlo approach, is the inclusion of a uniform magnetic field. 
Based on the mean-field description of the $\pi$-flux phase, we can speculate  
that the point-like Fermi {\it surface} at  zero field evolves to rings around centered around 
the zero field nodes. Since we are at 
a particle-hole symmetric point and that this symmetry is not broken under the inclusion of a 
magnetic field, the finite field Fermi surface is unstable towards magnetic ordering. Very much as in 
discussed in Ref. \cite{Milat04} this produces a 
field induced  transition to an magnetically ordered state.

Acknowledgments.  The calculations presented here were carried out on the IBM p690 cluster of 
the NIC  in J\"ulich. 
I would like to thank this institution for generous allocation of CPU time.  Part of this 
work has been carried out at the KITP, in the framework  of a program on Exotic order 
and Criticality in Quantum Matter.  
I have greatly  profited from discussions with 
M. Hermele, M. Imada. B. Marston, G. Misguich,   A. Paramekanti, S. Sachdev  and  J. Zaanen.
This research was supported in part by the National Science Foundation under Grant No. PHY99-07949.


\begin{thebibliography}{10}

\bibitem{Kugel82}
K.~I. Kugel' and D.~I. Khomskii, Sov. Phys. Usp. {\bf 25},  232  (1982).

\bibitem{Li88}
Y.~Q. Li, M. Ma, D.~N. Shi, and F.~C. Zhang, Phys. Rev. Lett. {\bf 81},  3527
  (1988).

\bibitem{Honerkamp04}
C. Honerkamp and W. Hofstetter, Phys. Rev. Lett.  (2004).

\bibitem{Assaad02a}
F.~F. Assaad, V. Rousseau, F. H\'ebert, M. Feldbacher, and G. Batrouni,
  Europhys. Lett. {\bf 63},  569  (2003).

\bibitem{Wu03}
C. Wu, J.~P. Hu, and S.~C. Zhang, Phys. Rev. Lett. {\bf 91},  186402  (2003).

\bibitem{Read89}
N. Read and S. Sachdev, Nucl. Phys. B {\bf 316},  609  (1989).

\bibitem{Santoro99}
G. Santoro, S. Sorella, L. Guidoni, A. Parola, and E. Tosatti, Phys. Rev. Lett.
  {\bf 83},  3065  (1999).

\bibitem{Harada03}
K. Harada, N. Kawashima, and M. Troyer, Phys. Rev. Lett. {\bf 90},  117203
  (2003).

\bibitem{Affleck88}
I. Affleck and J.~B. Marston, Phys. Rev. B {\bf 37},  3774  (1988).

\bibitem{Marston89}
J.~B. Marston and I. Affleck, Phys. Rev. B {\bf 39},  11538  (1989).

\bibitem{Dombre89}
T. Dombre and G. Kotliar, Phys. Rev. B {\bf 38},  855  (1989).

\bibitem{Capponi00}
S. Capponi and F.~F. Assaad, Phs. Rev. B {\bf 63},  155114  (2001).

\bibitem{Arovas88}
D. Arovas and A. Auerbach, Phs. Rev. B {\bf 38},  316  (1998).

\bibitem{Affleck88a}
I. Affleck, Z. Zou, T. Hsu, and P. Anderson, Phys. Rev. B {\bf 38},  745
  (1988).

\bibitem{Gross88}
C. Gross, Phys. Rev. B {\bf 38},  931  (1988).

\bibitem{Paramekanti04}
A. Paramekanti, M. Randeria, and N. Trivedi,
  http://xxx.lanl.gov/cond-mat/0405353  .

\bibitem{Hermele04}
M. Hermele, T. Senthil, M.~P.~A. Fisher, P.~A. Lee, N. Nagaosa, and X.~G. Wen,
  http://xxx.lanl.gov/cond-mat/0404751  .

\bibitem{Rantner02}
W. Rantner and X. Wen, Phys. Rev. B {\bf 66},  144501  (2002).

\bibitem{Ivanov02}
D. Ivanov and T. Senthil, Phys. Rev. B {\bf 66},  115111  (2002).

\bibitem{Misguich02}
G. Misguich, D. Serban, and V. Pasquier, Phys. Rev. Lett. {\bf 89},  137202
  (2002).

\bibitem{Milat04}
I. Milat, F. Assaad, and M. Sigrist, http://xxx.lanl.gov/cond-mat/0312450  .

\end{thebibliography}

\end{document}